\documentstyle[prd,aps,floats]{revtex}

\begin{document}

\draft
\input epsf
\twocolumn[\hsize\textwidth\columnwidth\hsize\csname
@twocolumnfalse\endcsname

\title{Remarks on Cosmic String Formation during Preheating 
                                   on Lattice Simulations} 

\author{S. Kasuya$^{1,2}$ and M. Kawasaki$^{1,3}$}
\address{${}^1$ Institute for Cosmic Ray Research, University of Tokyo,
  Tanashi, Tokyo 188-8502, Japan}
\address{${}^2$ Department of Physics, Ochanomizu University,
  Bunkyo-ku, Tokyo 112-8610, Japan}
\address{${}^3$ Research Center for the Early Universe, University 
  of Tokyo, Bunkyo-ku, Tokyo 113-0033, Japan}

\date{March 11, 1999}

\maketitle

\begin{abstract}
We reconsider the formation of (global) cosmic strings during and
after preheating by calculating the dynamics of a scalar field on
both two- and three-dimensional lattices. We have found that there
is little differences between the results in two and three
dimensions about the dynamics of fluctuations, at least, during
preheating. Practically, it is difficult to determine whether long 
cosmic strings which may affect the later evolution of the universe
could ever be produced from the results of simulations on
three-dimensional lattices with smaller box sizes than the horizon.
Therefore, using two-dimensional lattices with large box size, we have
found that cosmic strings with the breaking scale $\eta \sim 10^{16}
{\rm GeV}$ are produced for broad range of parameter space in $\eta$,
while for higher breaking scales  
($\eta \sim 3\times 10^{16} {\rm GeV}$), their production depends
crucially on the value of the breaking scale $\eta$ in our
simulations.   
\end{abstract}

\pacs{PACS numbers: 98.80.Cq, 11.27.+d
      \hspace{5cm} hep-ph/9903324}

%\newpage
\vskip2pc]

\setcounter{footnote}{1}
\renewcommand{\thefootnote}{\fnsymbol{footnote}}

One of the hot topics at the reheating stage after inflation is the
possibility of the formation of topological defects
\cite{KLS2,Tkachev,KK1,KK2,TKKL}. This phenomenon is due to large
nonthermal fluctuations during preheating \cite{KLS1,STB,Boyan,Yoshi}
and efficient rescattering \cite{KhTk1,KLS3} both of which are caused
by the Bose enhancement effects. 

At the preheating stage, very large nonthermal fluctuations are 
produced, $\langle \delta \phi^2 \rangle \sim c^2 M_p^2$ where 
$M_p$ is the Planck mass and $c=10^{-2}-10^{-3}$ \cite{KLS2,Tkachev}. 
These fluctuations make the shape of the effective potential of 
the field $\phi$ change to the type of the potential which has a 
minimum at the origin if the potential $V(\phi)$ is of spontaneous 
symmetry-breaking type. It may be regarded that the symmetry is 
restored \cite{KLS2,Tkachev,TKKL}. Later, when the amplitude of these 
fluctuations is redshifted away by the cosmic expansion, the symmetry 
is spontaneously broken and topological defects may be created. Thus, 
the mechanism for producing the topological defects seems to be 
somewhat similar to the Kibble mechanism in high temperature theory. 

For the order estimation of the critical value of the breaking scale 
where cosmic strings are not formed above that value, it is 
sufficient to see the amplitude of the nonthermal fluctuations 
produced during preheating: 
$\langle \delta \phi^2 \rangle^{1/2} \sim 10^{16} {\rm GeV}$ 
\cite{KLS2}. This order estimation is in good agreement with 
numerical estimations done in Refs.\cite{KK2,TKKL}. But numerical 
simulations on the lattices which follow the dynamics of the scalar 
fields reveals several new results, such as the effects of the 
rescattering \cite{KK2,TKKL} or the number of defects estimation
\cite{KK2}. 

In our previous paper \cite{KK2}, we investigated the dynamics of a
complex scalar field using two-dimensional lattice simulations, taking
into account that the box size is large enough to cover the horizon
size and that the lattice size is small enough to identify cosmic
strings safely, and concluded that (long) cosmic strings would not be
produced if the breaking scale $\eta$ was larger than 
$\eta \sim 3\times 10^{16} {\rm GeV}$. On the other hand, the authors
of Ref.~\cite{TKKL} showed the possibility that cosmic strings could
be formed even if $\eta=6\times 10^{16} {\rm GeV}$ using the 
three-dimensional lattice where the box size is smaller than the
horizon. In this paper, we will discuss both two- and
three-dimensional lattice simulations, and make certain that these
results are consistent, commenting on the limitations of both
simulations.

First we show that results from lattice simulations in two and three
dimensions are not different from each other when we study the
parametric resonance during preheating. To be concrete, let us
consider a complex scalar field with the effective potential:
\begin{equation}
    V(\Phi)=\frac{\lambda}{2}(|\Phi|^2-\eta^2)^2,
\end{equation}
where $\lambda$ is a small coupling constant. This model has a global
U(1) symmetry, and cosmic strings are formed when the symmetry is
spontaneously broken. What we have to do is to integrate the equation
of motion:
\begin{equation}
    \ddot{\Phi}+3H\dot{\Phi}-\frac{1}{a^2}\nabla^2\Phi
    +(|\Phi|^2-\eta^2)\Phi=0.
\end{equation}
For numerical simulations, it is convenient to use rescaled variables:
\begin{eqnarray}
  \label{re-sf}
    a(\tau)d\tau & = & \sqrt{\lambda}\Phi_0 a(0)dt, \\
    \varphi & = & \frac{\Phi a(\tau)}{\Phi_0 a(0)}, \\
    \xi & = & \sqrt{\lambda}\Phi_0 a(0)x,
\end{eqnarray}
where $\Phi_0\equiv|\Phi(0)|$. Setting $a(0)=1$, we obtain 
\begin{equation}
  \label{eom-2}
    \varphi^{\prime\prime} - \frac{a^{\prime\prime}}{a}\varphi
             - \nabla^2_{\xi}\varphi 
             + ( |\varphi|^2 - \tilde{\eta}^2 a^2 )\varphi =0,
\end{equation}
where $\tilde{\eta}\equiv\eta/\Phi_0$ and the prime denotes
differentiation with respect to $\tau$. The second term of the
left-handed side can be omitted since the energy density of the
universe behaves like radiation at the early times and also the scale
factor becomes very large later. In this case, the rescaled Hubble
parameter,$h(\tau)\equiv H(\tau)/\sqrt{\lambda}\Phi_0$, and the scale
factor $a(\tau)$ become \cite{KK2}
\begin{equation}
    h(\tau) = \frac{\sqrt{2}}{3}a^{-2}(\tau),
\end{equation}
and
\begin{equation}
  \label{sf}
    a(\tau) = \frac{\sqrt{2}}{3}\tau + 1,
\end{equation}
respectively, when $\Phi$ is assumed to be an inflaton (even if $\Phi$ 
is not an inflaton, results are the same as in the case of rescaling
the breaking scale in an appropriate way, see Ref.~\cite{KK2}) 
and we set $a(0)=1$. For the initial conditions we take 
\begin{eqnarray}
    x \equiv {\rm Re}\varphi(0) & = &
                               1 + \delta x({\bf x}), \\
    y \equiv {\rm Im}\varphi(0) & = &
                               \delta y({\bf x}),         
\end{eqnarray}
where the homogeneous part comes from its definition (we call
$x$ direction for real direction and $y$ for imaginary), 
and $\delta x,y({\bf x})$ is a small random variable of 
${\cal O}(10^{-7})$ representing the fluctuations. We also set small
random values for velocities. 

Since the physical length and the horizon grow proportional to $a$ and 
$a^2$, respectively, the rescaled horizon grows proportional to
$a$. The initial length of the horizon is 
$\ell_h(0)=3/\sqrt{2} \approx 2.12$. Therefore, the rescaled horizon
size grows as $\ell_h(\tau)=\ell_h(0)a(\tau)$. It is thus better to
take the box size larger than $\ell(\tau_{end})$ where $\tau_{end}$ is 
the time at the end of the simulation. This is because those strings
whose lengths are shorter than the horizon scale will be in the form
of loops, which will shrink and disappear very soon. Only longer
strings than the horizon will survive to affect the later evolution of
the universe. On the other hand, the
width of the topological defect is $(\sqrt{\lambda}\eta)^{-1}$ 
which corresponds to $(\tilde{\eta}a(\tau))^{-1}$ in the rescaled
units. Since it decreases with time, one lattice length should be at
least comparable with the defect width at the end of the calculation. 

\begin{figure}[t!]
\centering
\hspace*{-5.5mm}
\leavevmode\epsfysize=6.5cm \epsfbox{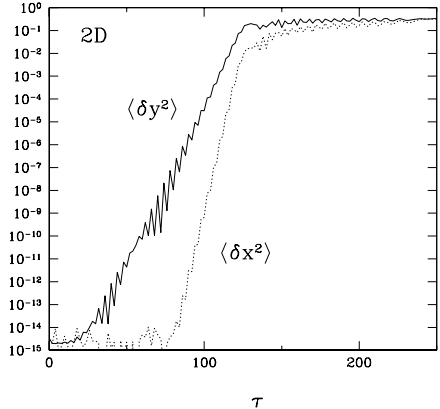}\\[3mm]
\hspace*{-5.5mm}
\leavevmode\epsfysize=6.5cm \epsfbox{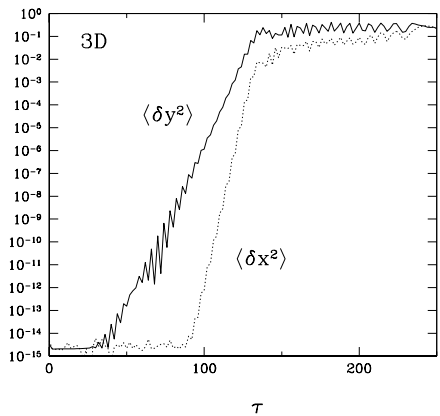}\\[3mm]
\caption{\label{eta316} 
Evolution of the fluctuations for $\eta=3\times10^{16}{\rm GeV}$ on 
two- (top panel) and three- (bottom panel) dimensional lattices. The 
solid and dotted line denote the fluctuations in $y$ and $x$ 
directions, respectively. Notice that the data used on the
two-dimensional lattices is the same used in Fig.4 of Ref.[4]. }
\end{figure}

Leaving these facts aside, let us just compare the evolution of
fluctuations on two-dimensional lattices with three-dimensional ones.
Here we take $128^3$ three-dimensional lattices and $4096^2$ 
two-dimensional lattices with the lattice size $\Delta\xi=0.3$ for
both cases. We find no difference between both growth 
exponents (see Fig.~\ref{eta316}). Notice that the data used on the 
two-dimensional lattice (top panel) is the same used in Fig.4 of 
Ref.~\cite{KK2}, where it was plotted linearly instead of 
logarithmically as in Fig.~\ref{eta316}.
\footnote{%
 The authors of Ref.~\cite{TKKL} might be misleaded by this point,
 since they claimed that the growth exponent of the fluctuation in the
 $x$ direction is much larger in three-dimensional lattices than that 
 in the two-dimensional cases of our previous results in
 Ref.~\cite{KK2}.} 
Moreover, we obtain very similar results in two- and three-dimensional 
lattices, at least, on the effects of parametric resonance during
preheating.

\begin{figure}[t!]
\centering
\hspace*{-7mm}
\vspace*{-4mm}
\leavevmode\epsfysize=5.5cm \epsfbox{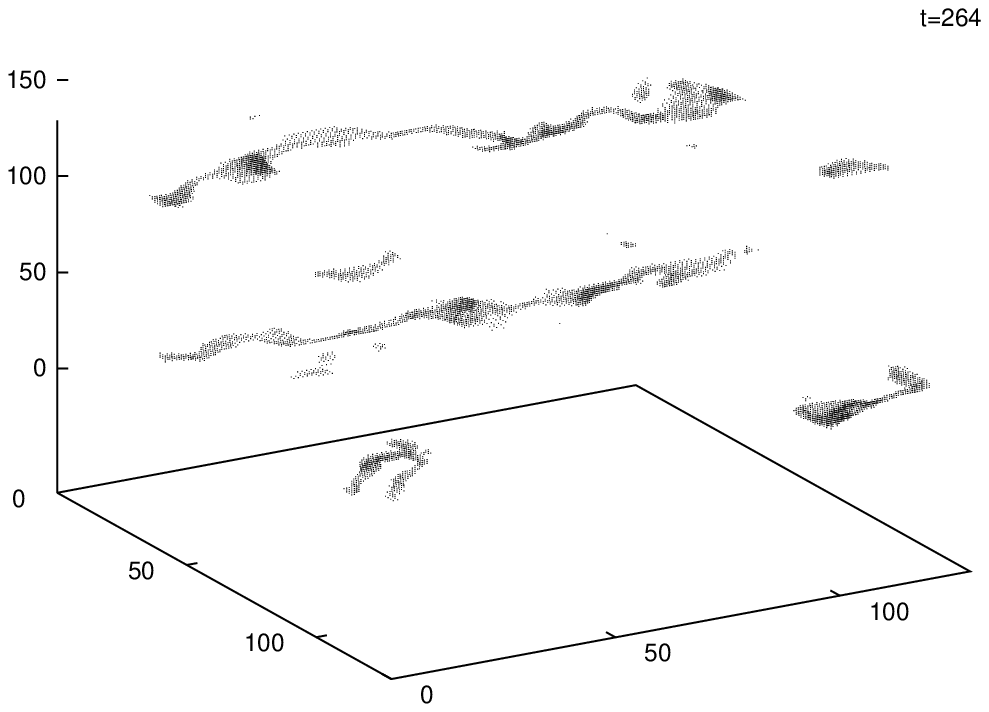}\\[3mm]
\hspace*{-7mm}
\vspace*{-4mm}
\leavevmode\epsfysize=5.5cm \epsfbox{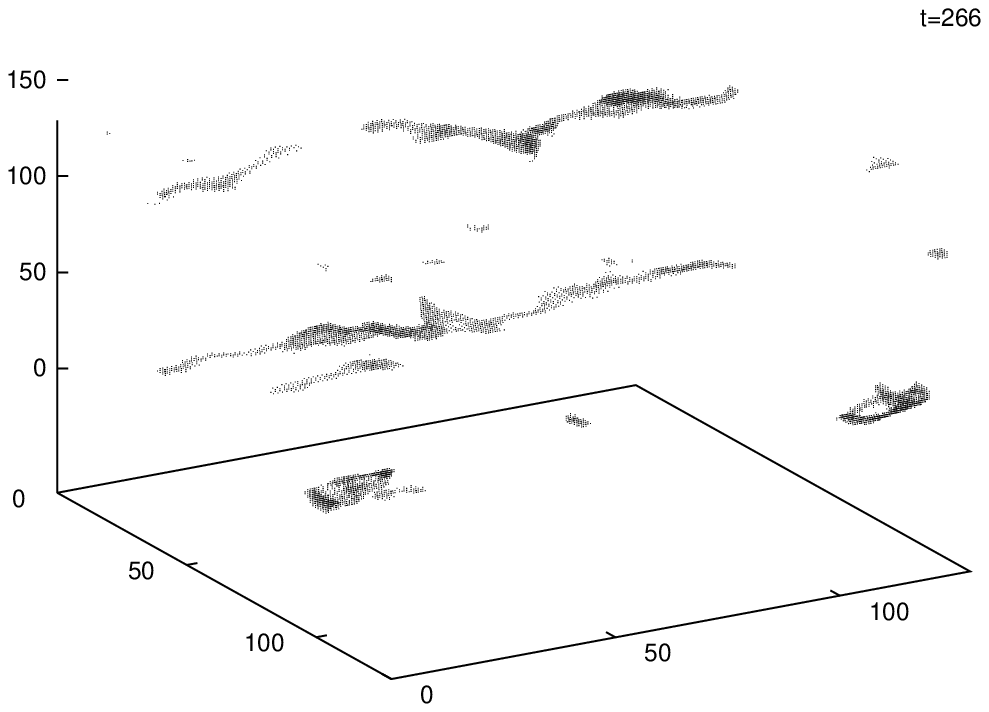}\\[3mm]
\hspace*{-7mm}
\vspace*{-4mm}
\leavevmode\epsfysize=5.5cm \epsfbox{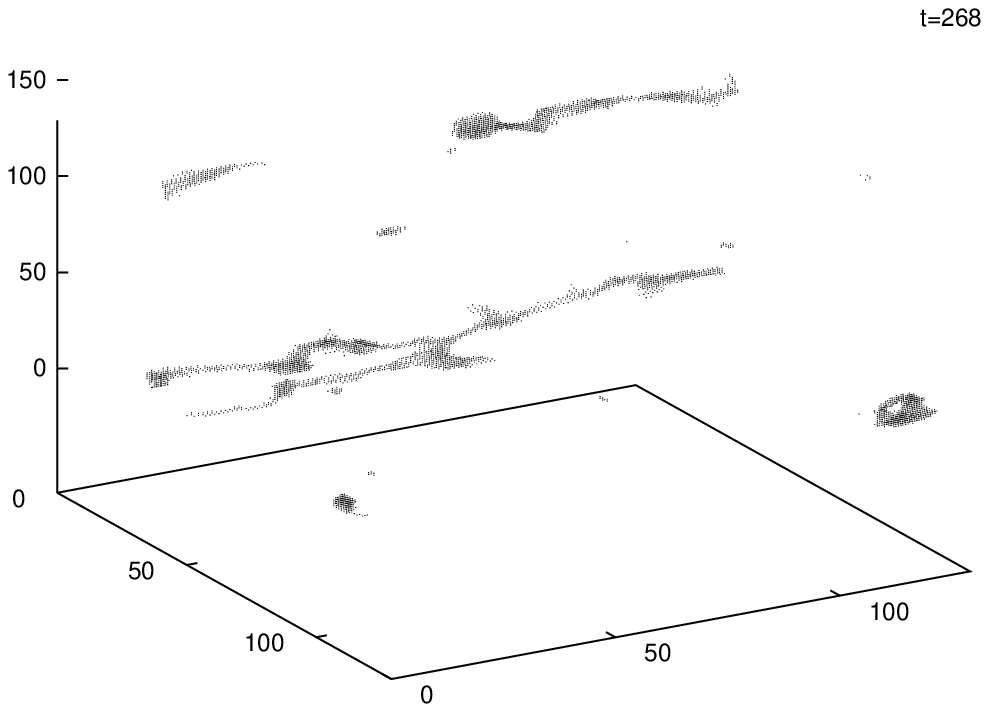}\\[3mm]
\hspace*{-7mm}
\vspace*{-4mm}
\leavevmode\epsfysize=5.5cm \epsfbox{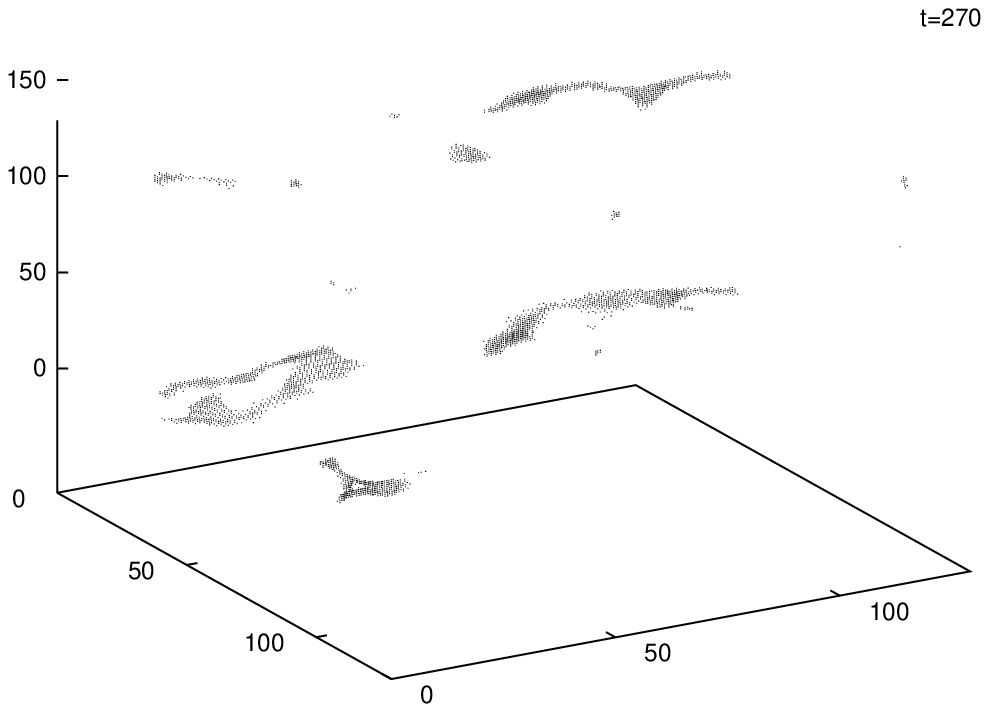}\\[3mm]
\caption{\label{3d-b3} 
Evolution of cosmic strings for $\eta=3.08\times10^{16}$ GeV. We plot 
from $\tau=264$ (top) to $\tau=270$ (bottom).}
\end{figure}

Since small loop strings will shrink and disappear very soon, and
cause no influence on the universe, we are interested only in 
infinitely long strings. In three-dimensional lattices, they can be
considered as those strings that penetrate through the box of the
lattice (come into the box from one side and go out to the other side)
and survive until the later time. Actually, we have found temporary 
formation of cosmic strings at any breaking scale, which confirms the
result of Ref.~\cite{TKKL}. In most values of the breaking scale,
however, these strings are in the form of small loops, and disappear
very quickly. We have also found strings longer than the box size. In
Fig.~\ref{3d-b3}, we take the breaking scale 
$\eta=3.08\times10^{16}{\rm GeV}$, the lattice size $\Delta\xi=0.3$,
and integrate the equation of motion until $\tau=280$. If we
integrated until the time before $\tau=265$, we would conclude that
cosmic strings were formed for this breaking scale. However, as we can
see in  Fig.~\ref{3d-b3}, these long strings feel the attractive force
from each other, and they form into loops beyond the box size, which
is much smaller than the horizon scale:  
$N\Delta\xi = 38.4 \ll \ell_h(\tau=280)\approx 282$, where $N$ is the
number of the lattice.

We agree with the authors of Ref.~\cite{TKKL} that longer strings tend
to be created when the velocity of the real part of the oscillating
homogeneous mode $x$ becomes almost zero at the moment when it passes
through the origin of the effective potential (in their words, when
the moment of the symmetry breaking nearly coincides with the moment
when $\langle\phi_1\rangle(=x)$ passes through zero \cite{TKKL}). They
argued that it leads to the fact that the long string formation is a
non-monotonic function of the breaking scale \cite{TKKL}, which we
confirm to some extent. However, in more than one hundred runs of our
simulations, for almost all the breaking scales higher than
$3\times10^{16}{\rm GeV}$, we have not found long strings stretched
out beyond the box size except for two breaking scales: 
\footnote{%
  We have recently found long strings formed at 
  $3.01\times 10^{16} {\rm GeV}$, but they are very unstable in the
  sense that strings in the form of loops develop into the form of
  long strings, and later they again deform into loops and disappear. 
  Notice that the range of the breaking scale where long strings are
  formed is also very narrow at this scale as in the above  
  two scales. Actually, these scales correspond to the circumstances
  that the velocity of the real part of the oscillating homogeneous
  mode $x$ becomes almost (but not exactly) zero at the moment when it
  passes through the origin of the effective potential, so there may
  be more breaking scales where long cosmic strings are formed.}
$\eta\simeq3.08\times10^{16}{\rm GeV}$ and 
$\eta\simeq3.16\times10^{16}{\rm GeV}$ on the $128^3$ lattices with
$\Delta\xi=0.3$ (Actually, long strings cannot be found at these
scale, if another initial configurations of initial fluctuations are
used). Even in these scales, long strings deform into loops and shrink
and disappear very soon. 

Therefore, we cannot make any definite conclusion 
on the formation of cosmic strings in the sense that we cannot tell
whether or not they may affect the evolution of the universe. 
This is because the box size cannot be taken to be larger than the
horizon scale in three-dimensional lattices. We have thus studied the
formation of cosmic strings in two-dimensional lattices in the
previous paper \cite{KK2}, abandoning one dimension in space because
of the lack of memory capacity of computers.  

\begin{figure}[t!]
\centering
\hspace*{-7mm}
\leavevmode\epsfysize=8cm \epsfbox{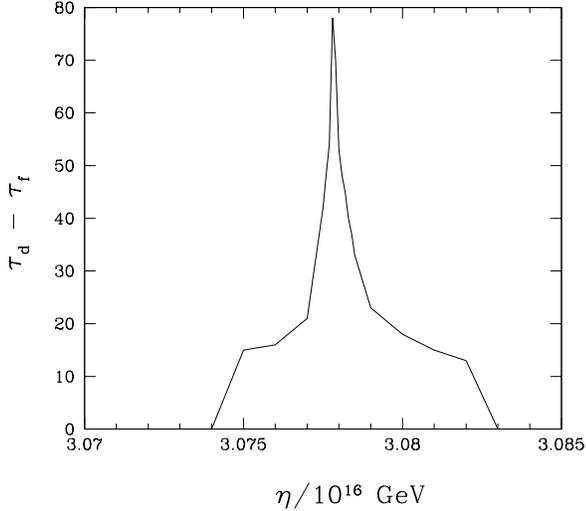}\\[2mm]
\caption[fig-ho]{\label{fig-1} 
Dependence of the lifetime ($\tau_d-\tau_f$) when long strings 
penetrate the box of the lattice on the breaking scale $\eta$ around
$\eta=3.078\times10^{16} {\rm GeV} $. Here $\tau_d$ is the destruction 
time when a long string deforms into a loop, and $\tau_f \approx 250$
is the formation time when a long string is first formed.
Notice that only loop strings can be formed outside the narrow strip
of the breaking scale ($\eta<3.075\times10^{16} {\rm GeV} $, 
$3.082\times10^{16} {\rm GeV} < \eta$).}
\end{figure}

\begin{figure}[t!]
\centering
\hspace*{-7mm}
\leavevmode\epsfysize=8cm \epsfbox{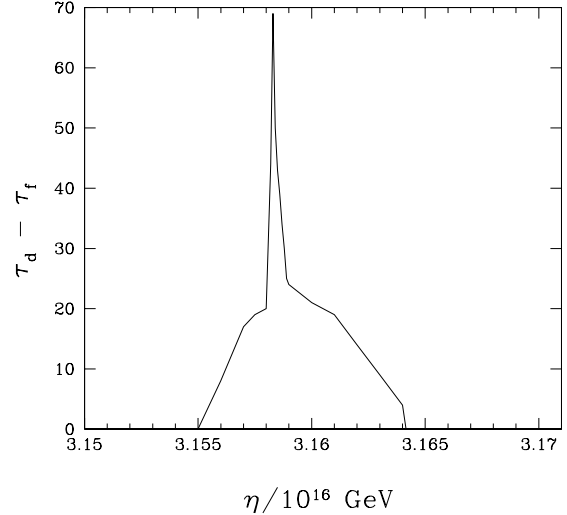}\\[2mm]
\caption[fig-ho]{\label{fig-2}
Dependence of the lifetime ($\tau_d-\tau_f$) when long strings
penetrate the box of the lattice on the breaking scale $\eta$ around
$\eta=3.158\times10^{16} {\rm GeV}$. Notice that the formation time
is $\tau_f \approx 250$, and only loop strings can be formed outside
the narrow strip of the breaking scale 
($\eta<3.156\times10^{16} {\rm GeV}$, 
$3.164\times10^{16} {\rm GeV} < \eta$).}
\end{figure}

Nevertheless, we can extract useful information from the results on
three-dimensional lattices. If we take the criterion of the formation
of (infinitely) long strings as the cases when those strings stretched 
across the box size have been once formed, for the sake of the
conservative discussion, we have found that the corresponding cases
are in the only very narrow region of the parameter space of the
breaking scale for the fixed initial conditions:
\footnote{%
 For other initial conditions, narrow ranges of the breaking scale
 where long cosmic strings are produced appear in the different
 scales, and they are also very narrow.}
$\Delta\eta/\eta \simeq 3\times 10^{-3}$. See Figs.~\ref{fig-1} and 
\ref{fig-2}. Figures \ref{fig-1} and \ref{fig-2} show the lifetime
($\tau_d -\tau_f$) with respect to the breaking scale $\eta$ for 
$\sim 3.08\times10^{16} {\rm GeV}$ and 
$\sim 3.16\times10^{16} {\rm GeV}$, respectively. We can see that the
size of the narrow regions are both 
$\Delta\eta \simeq 0.009\times 10^{16}$ GeV. Here $\tau_d$ is the time
when the long strings destruct into loops, and $\tau_f$ is the
formation time for those strings. 

However, this criterion may not be so sufficient, since those strings
stretched across the lattice box will deform into loops and disappear 
very soon. We can take other criterion, which may reflect the
idea that it is important to consider a long cosmic string. When we
observe the lifetime ($\tau_d -\tau_f$), we find that long strings
survive a few times longer at a certain range of the breaking scale,
as shown in Figs.~\ref{fig-1} and \ref{fig-2}. We can expect that long 
strings that stretch beyond the horizon size will be formed only
within such very narrow regions. In this case, therefore, the breaking
scale should be in the very narrow ranges 
($\Delta\eta/\eta\simeq 10^{-4}$) in order for (infinitely) long
strings to be formed for the fixed initial conditions in our
simulations (These features can be also seen on the lattices with
larger box size: $N=200$). In other words, the long string formation
is very sensitive to the breaking scale.

These results can be understood as follows. If there is no gradient
force, the dynamics of the field is determined only by the homogeneous
mode. But the initial values of the field at each point on the lattice 
differs from each other by ${\cal O}(10^{-7})$, we naively expect that
cosmic strings are formed only in the very narrow ranges of the
breaking scale of ${\cal O}(10^{-7})$. Owing to the preheating stage,
fluctuations become large, so that the degree of the narrow ranges is
somewhat relaxed. However, it seems that the full development of
rescattering does not occur yet for 
$\eta\sim 3\times10^{16} {\rm GeV}$, as we mentioned in
Ref.~\cite{KK2}. We will see later that the fundamentally different
feature appears for the lower breaking scales. 

\begin{figure}[t!]
\centering
\hspace*{-7mm}
\leavevmode\epsfysize=8cm \epsfbox{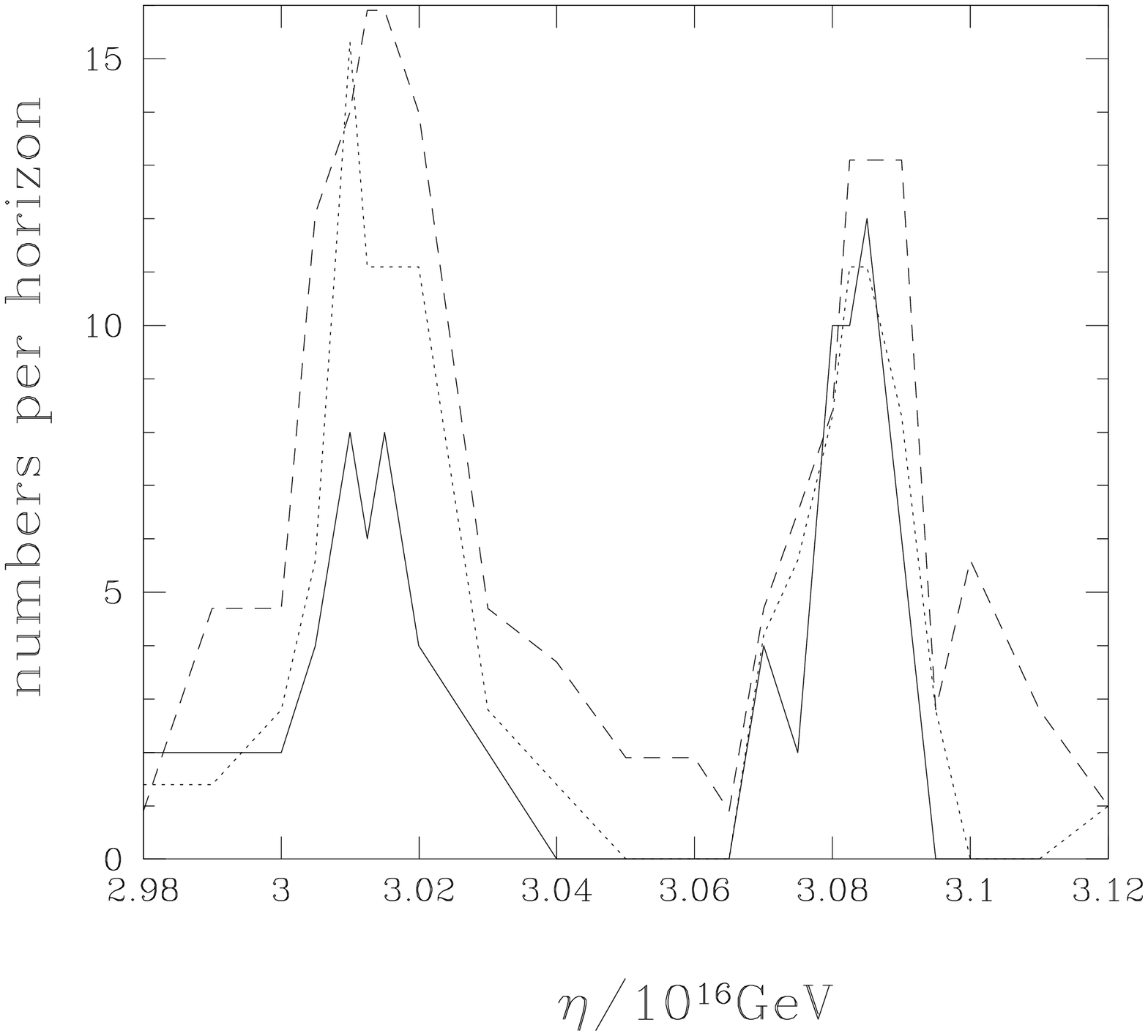}\\[2mm]
\caption[fig-ho]{\label{fig-5} 
Dependence of the number of strings per horizon on the breaking scale
$\eta$ around $\eta=3\times10^{16} {\rm GeV}$ at $\tau=410$ (dashed), 
$\tau=500$ (dotted), and $\tau=600$ (solid), in two-dimensional
simulations.} 
\end{figure}

As mentioned above, it is difficult to tell whether long cosmic
strings are formed on three-dimensional lattices with a smaller box
size than the horizon volume, and this is why we have calculated in
two dimensions, in order to make definite conclusions using both two
and three-dimensional simulations complementarily. Similar results are 
found in two-dimensional lattices. In the two-dimensional case, we
cannot distinguish long strings from loops, because all the strings
are assumed to be infinitely stretched along the $z$ direction. 
Instead, we observe the number of cosmic strings in the horizon size
at each time. The physically meaningful criterion for the formation of
long cosmic strings is whether the number of strings per horizon
remain (almost) constant or not as time goes on. Even if we cannot
distinguish very long strings from loop ones, we can regard very
nearly located string-antistring pair in two dimensions as a small
loop string in three dimensions. Actually, they annihilate very soon,
similar to small loop strings which will shrink and disappear very
quickly. Thus, when we observe the number of strings as time goes on,
it will remain at a certain value if long strings are formed, since
loop strings (string-antistring pairs in terms of two dimensions) will 
disappear and very long strings (isolated strings in terms of two
dimensions) will remain. Notice that we include both strings and
antistrings in the numbers. As a result, we find that it depends on
the breaking scale very crucially at the scale 
$\eta\sim 3\times10^{16}{\rm GeV}$, as is seen in Fig.~\ref{fig-5}. At
the scales $\eta\sim 3.02 \times10^{16}{\rm GeV}$ and 
$\eta\sim 3.09 \times10^{16}{\rm GeV}$, a dozen cosmic strings are
formed, and the numbers do not decrease so much. On the other hand, at
the other scales, the numbers of strings decreases, and finally, we
find no strings at all. Thus, the value of the breaking scale has to
be in the very narrow ranges such as the degree of 
$\Delta\eta/\eta\simeq10^{-2}$, where $\Delta\eta$ is defined as the
breaking scales at which the number of strings per horizon remains at
a certain value as time goes on. This implies, together with the
results in three dimensions, that long string formation is very
sensitive to the breaking scale at these scales.

\begin{figure}[t!]
\centering
\hspace*{-7mm}
\leavevmode\epsfysize=8cm \epsfbox{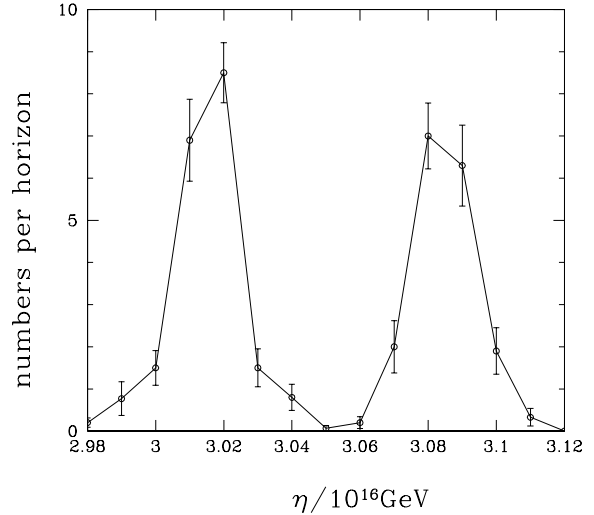}\\[2mm]
\caption[fig-ho]{\label{fig-6} 
Average numbers of strings per horizon at $\tau=600$ over $30$
realizations for each breaking scale $\eta$ around
$\eta=3\times10^{16} {\rm GeV}$ in two-dimensional simulations. } 
\end{figure}

Moreover, we have simulated $30$ realizations of initial conditions
for fluctuations for each breaking scale, and find that the
dependence of the average numbers of strings per horizon on the
breaking scales, shown in Fig.~\ref{fig-6}, coincides with the above
particular one of Fig.~\ref{fig-5}. This not only confirms the crucial 
sensitiveness to the breaking scale, but also implies that the main
factor which determines whether long strings are produced is the value 
of the breaking scale. Initial fluctuations do not affect the dynamics
of the scalar field so much at $\eta \sim 3\times 10^{16}$
GeV~\cite{KK1}.  

\begin{figure}[t!]
\centering
\hspace*{-7mm}
\leavevmode\epsfysize=8cm \epsfbox{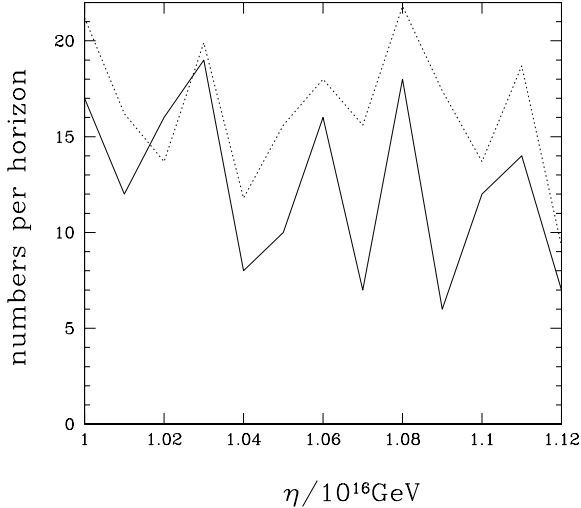}\\[2mm]
\caption[fig-ho]{\label{fig-7} 
Dependence of the number of strings per horizon on the breaking scale
$\eta$ around $\eta=10^{16} {\rm GeV} $ at $\tau=1420$ (dotted), and 
$\tau=1800$ (solid) in two-dimensional simulations.} 
\end{figure}

On the contrary to the cases with somewhat higher breaking scale such
as $\eta\sim 3 \times10^{16}{\rm GeV}$, many cosmic strings are
formed at $\eta\sim 10^{16}{\rm GeV}$ as is seen in
Fig.~\ref{fig-7}. We have found more than a dozen strings per
horizon size at any value of the breaking scales near 
$\eta\sim 10^{16}{\rm GeV}$, and the numbers do not decrease so much. 
We thus conclude that the formation of cosmic strings occurs in the
very broad region of the breaking scale at $\eta\sim 10^{16}$ GeV. 

\begin{figure}[t!]
\centering
\hspace*{-5.4mm}
\leavevmode\epsfysize=6.6cm \epsfbox{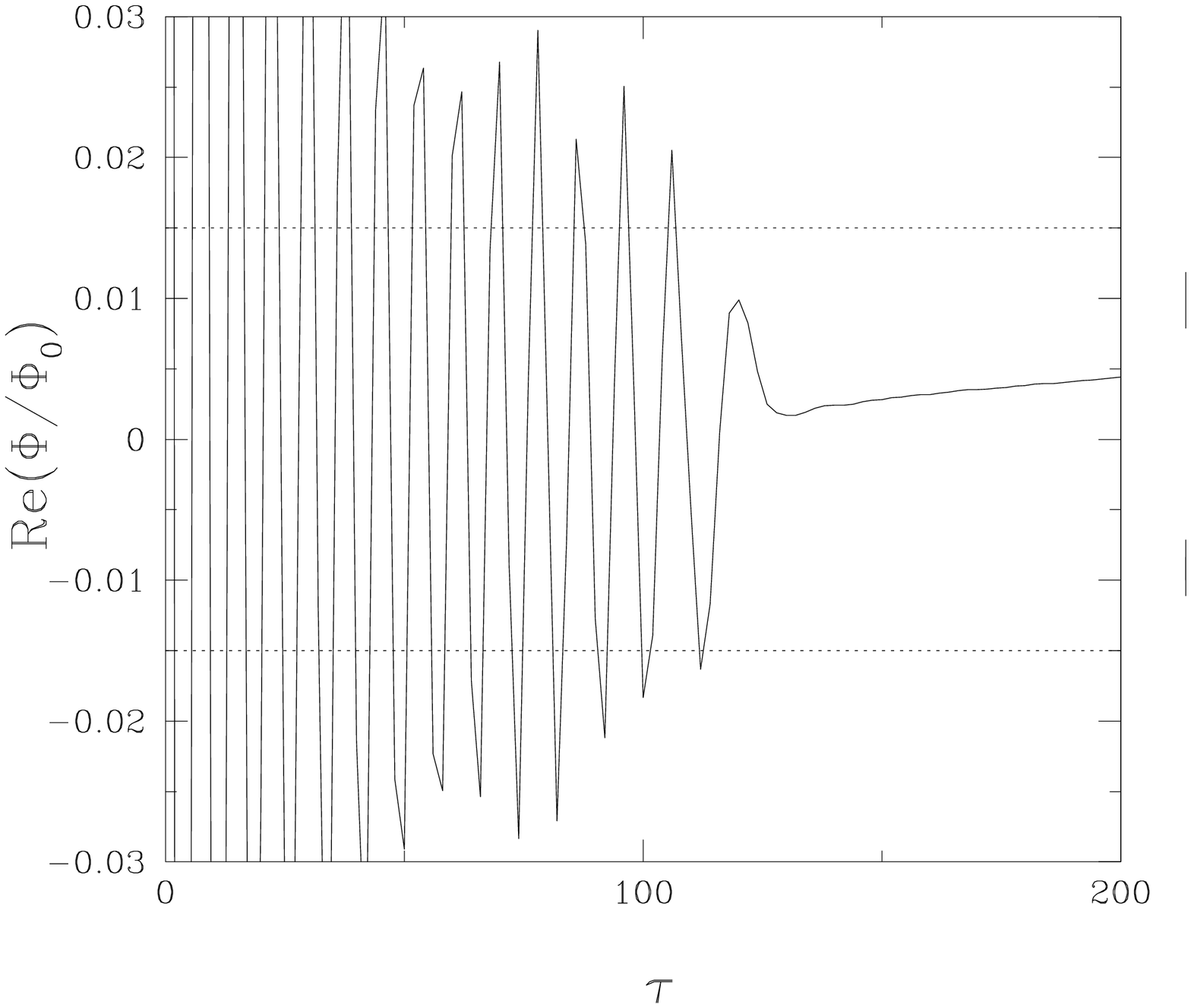}\\[3.5mm]
\hspace*{-5.4mm}
\leavevmode\epsfysize=6.6cm \epsfbox{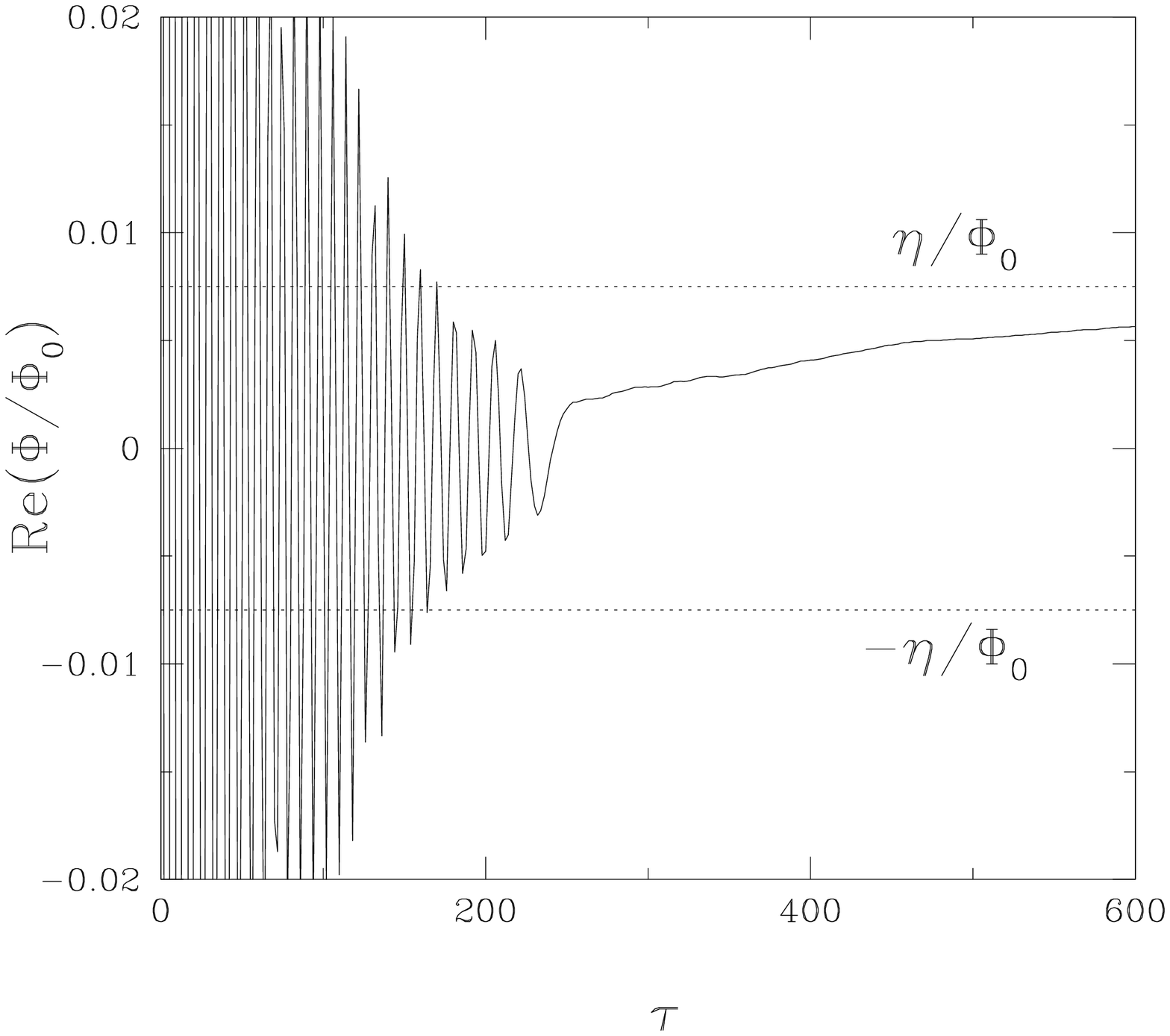}\\[3.5mm]
\hspace*{-5.4mm}
\leavevmode\epsfysize=6.6cm \epsfbox{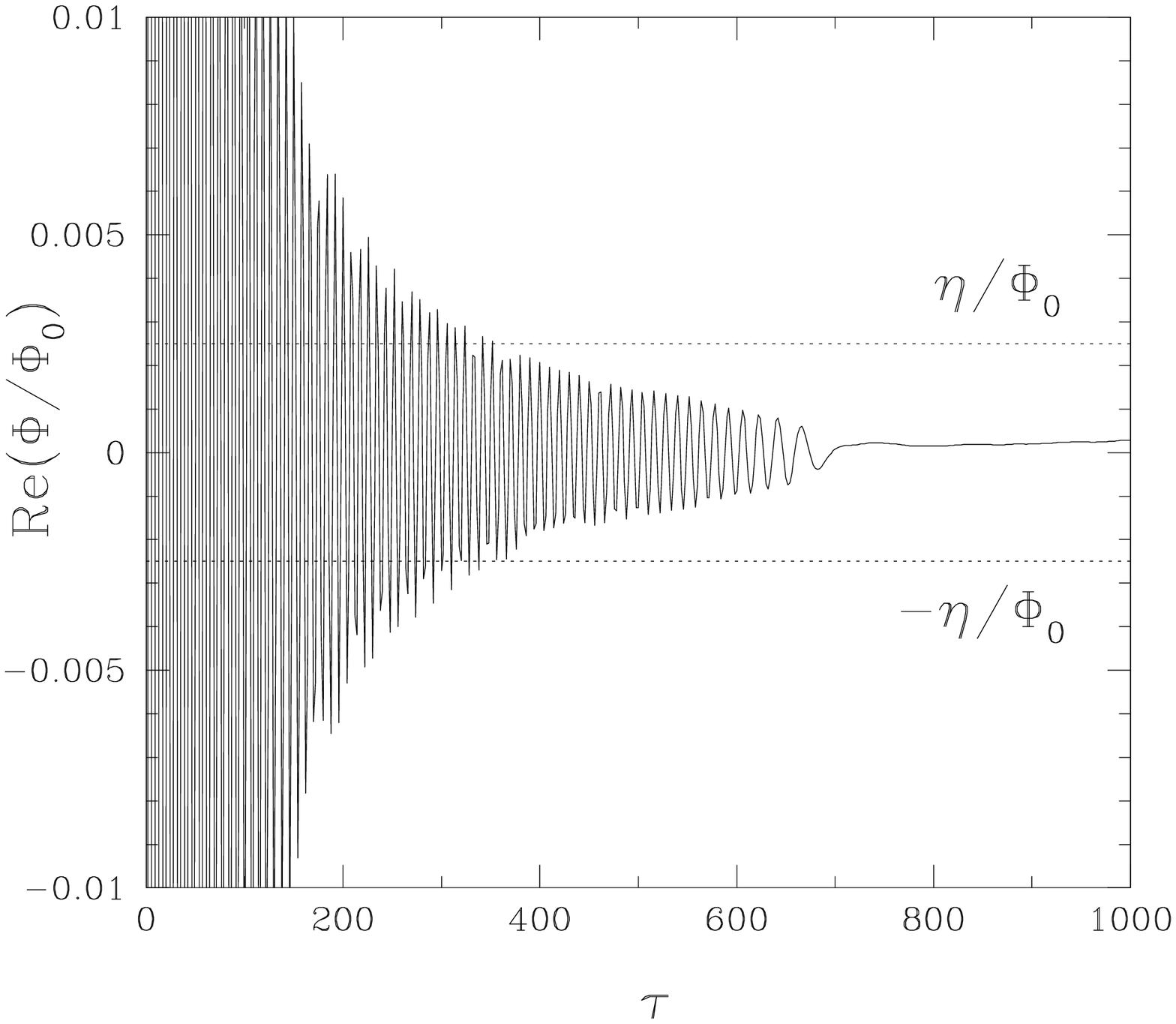}\\[3.5mm]
\caption{\label{sym-res} 
Evolution of the $\Phi$-field for various breaking scales. 
$\eta = 6\times10^{16}{\rm GeV}$, $\eta = 3\times10^{16}{\rm GeV}$,
and $\eta = 10^{16}{\rm GeV}$ from the top to the bottom. Here the
amplitudes are plotted in the physical unit, not in the rescaled one.}
\end{figure}

Finally, we should comment on the relation between the symmetry
restoration and the topological defect (cosmic string) formation. To
know whether the symmetry is restored or not is a difficult task to
do, and its methods are somewhat uncertain. The authors of
Ref.~\cite{TKKL} argued it in terms of the shape of the effective
potential for the scalar field. If the potential has a minimum at the
origin, the symmetry is restored. We may make sure that the potential
has a minimum at the origin in the following way, as done in
Ref.~\cite{TKKL}. The field $\phi$ (radial direction of $\Phi$)
will oscillate around the origin in the potential of the form
$(\phi^2-\eta^2)^2$ when its amplitude is larger than
$\sqrt{2}\eta$. But, as we can see in Fig.~\ref{sym-res}, $\phi$ is
oscillating around the origin even when its amplitude becomes smaller
than the breaking scale $\eta$. It is thus very clear that the
effective potential should have a minimum at the origin. We can see
this phenomena in all cases in Fig.~\ref{sym-res}. However, in the
above, we actually see that topological defects are produced only when
the breaking scale is $\eta=10^{16}{\rm GeV}$, not in the cases of 
$\eta = 3\times10^{16}{\rm GeV}$ and 
$\eta = 6\times10^{16}{\rm GeV}$. Therefore, the symmetry must be 
fully restored only in the case of $\eta=10^{16}{\rm GeV}$, where
rescatterings play a crucial role for that. 

Then what is the relation between the symmetry restoration and defect
production? We conclude that defect formation {\it is} the signal of
the full restoration of symmetry. We thus discriminate between
symmetry restoration and the shape of the effective potential which
has a minimum at the origin. In other words, you cannot tell if the
symmetry is restored only by observing the shape of the potential. 

In conclusion, we have reconsidered the formation of (global) cosmic
strings during and after preheating by calculating the dynamics of the
scalar field on both two- and three-dimensional lattices, and
confirmed the results of both Ref.~\cite{TKKL} and our previous ones
\cite{KK2}. We have found that there are little differences between
the results in two and three dimensions, at least, at the preheating
stage. It is obvious that phase-space volume is larger in the three
dimensions than two, and this effect might affect somehow in the
rescattering stage, but we expect that it seems subdominant when we
obverse our numerical simulations. 

Practically, it is difficult to determine whether long cosmic strings
which may affect the later evolution of the universe could ever be
produced from the results of simulations on three-dimensional
lattices, since they will deform into large loops and disappear very
soon because of the small box size of the lattices. Moreover, we have
found that cosmic strings with a higher breaking scale than 
$3\times 10^{16} {\rm GeV}$ could only be produced in the very
narrow ranges of the breaking scale in our simulations. 
In Ref.~\cite{TKKL}, they referred to this formation of cosmic string
as the formation of a nonmonotonic function of the breaking scale. We
confirm this result and, in addition, find that the
formation of long cosmic strings occurs in very small parameter space 
of the breaking scale $\eta$. In other words, it is very sensitive to
the value of the breaking scale. In two-dimensional simulations, long
strings and loops can be distinguished to some extent, even though
it is difficult, by observing the evolution of the number of strings
per horizon, and we have found similar phenomenon that long strings
are produced only within very narrow range of the breaking scale
around  $\eta \sim 3\times 10^{16} {\rm GeV}$ for fixed initial
conditions. On the contrary, they are produced for a wide range of the 
breaking scale when $\eta \sim 10^{16} {\rm GeV}$.   

Cosmologically, we are interested only in (infinitely) long cosmic
strings stretched beyond the horizon. Those strings with breaking scale 
$\eta \lesssim 10^{16}$ GeV are naturally formed after preheating,
since they are produced independent to the actual values of the
breaking scale. In other words, they are produced and their numbers
remain constant in every horizon volume. 

On the other hand, when the breaking scale is larger 
($\eta \sim 3\times 10^{16}$ GeV), it is very difficult to connect
our results directly to the actual probability of the string
formation. What we have found is that the formation of a cosmic string
with $\eta \sim 3\times 10^{16}$ GeV depends crucially on the breaking 
scale. As mentioned, since initial fluctuations do not affect whether
long strings are formed so much, we can tell how many long strings are
produced, if the value of the breaking scale and the initial condition 
for the homogeneous mode are fixed, which is, in general, determined
if the inflation model is specified.

S.K. and M.K. would like to thank A. Linde for helpful comments.
S.K. is also grateful to L. Kofman and I. Tkachev for useful
discussions. M.K. is supported in part by the Grant-in-Aid, Priority
Area ``Supersymmetry and Unified Theory of Elementary
Particles''($\#707$).


\begin{references}
%%
\bibitem{KLS2} L. Kofman, A.D. Linde, and A. Strarobinsky,
    Phys. Rev. Lett. {\bf 76}, 1011 (1996).
%%
\bibitem{Tkachev} 
    I.I. Tkachev,
    Phys. Lett. {\bf B376}, 35 (1996).
%%
\bibitem{KK1} S. Kasuya and M. Kawasaki,
    Phys. Rev. {\bf D56}, 7597 (1997).
%%
\bibitem{KK2} S. Kasuya and M. Kawasaki,
    Phys. Rev. {\bf D58}, 083516 (1998).
%%
\bibitem{TKKL} I. Tkachev, S. Khlebnikov, L. Kofman, and A. Linde,
    Phys. Lett. {\bf B440}, 262 (1998). 
%%
\bibitem{KLS1} L. Kofman, A.D. Linde, and A. Strarobinsky,
    Phys. Rev. Lett. {\bf 73}, 3195 (1994).
%%
\bibitem{STB} Y. Shtanov, J. Traschen, and R. Brandenberger,
    Phys. Rev. {\bf D51}, 5438 (1995).
%%
\bibitem{Boyan} 
    D. Boyanovsky, H.J.de Vega, R. Holman, D.-S. Lee, and A. Singh,
    Phys. Rev. {\bf D51},4419 (1995).
%%
\bibitem{Yoshi} M. Yoshimura,
    Prog. Theor. Phys. {\bf 94}, 873 (1995).
%%
\bibitem{KhTk1} 
    S.Yu. Khlebnikov and I.I. Tkachev,
    Phys. Rev. Lett. {\bf 77}, 219 (1996).
%%
\bibitem{KLS3} L. Kofman, A.D. Linde, and A. Strarobinsky,
    Phys. Rev. {\bf D56}, 3258 (1997).

\end{references}
\end{document}